\documentclass[fleqn,usenatbib]{mnras}

\usepackage{newtxtext,newtxmath}

\usepackage[T1]{fontenc}

\DeclareRobustCommand{\VAN}[3]{#2}
\let\VANthebibliography\thebibliography
\def\thebibliography{\DeclareRobustCommand{\VAN}[3]{##3}\VANthebibliography}

\usepackage{graphicx}	
\usepackage{amsmath}	

\title[NSMs as the Dominant Contributor to Heavy Elements]{Neutron Star Mergers as the Dominant Contributor to the Production of Heavy $r$-Process Elements}

\author[M.-H. Chen et al.]{Meng-Hua Chen$^{1}$, Li-Xin Li$^{1}$\thanks{E-mail: lxl@pku.edu.cn}, Qiu-Hong Chen$^{2}$, Rui-Chong Hu$^{2}$, and En-Wei Liang$^{2}$\thanks{E-mail: lew@gxu.edu.cn}
\\
$^{1}$Kavli Institute for Astronomy and Astrophysics, Peking University, Beijing 100871, China\\
$^{2}$Guangxi Key Laboratory for Relativistic Astrophysics, School of Physical Science and Technology, Guangxi University, Nanning 530004, China\\
}

\date{Accepted XXX. Received YYY; in original form ZZZ}

\pubyear{2024}

\begin{document}
\label{firstpage}
\pagerange{\pageref{firstpage}--\pageref{lastpage}}
\maketitle

\begin{abstract}
The discovery of the radioactively powered kilonova AT2017gfo, associated with the short-duration gamma-ray burst GRB 170817A and the gravitational wave source GW170817, has provided the first direct evidence supporting binary neutron star mergers as crucial astrophysical sites for the synthesis of heavy elements beyond iron through $r$-process nucleosysthesis in the universe. However, recent identifications of kilonovae following long-duration gamma-ray bursts, such as GRB 211211A and GRB 230307A, has sparked discussions about the potential of neutron star-white dwarf mergers to also produce neutron-rich ejecta and contribute to the production of heavy $r$-process elements.
In this work, we estimate the contribution of binary neutron star mergers to the total mass of $r$-process elements in the Milky Way and investigate the possibility of neutron star-white dwarf mergers as alternative astrophysical sites for $r$-process nucleosynthesis through an analysis of the total mass of the $r$-process elements in the Milky Way. Our results reveal that binary neutron star mergers can sufficiently account for the Galactic heavy $r$-process elements, suggesting that these events are the dominant contributor to the production of heavy $r$-process elements in the Milky Way. Considering the total mass of $r$-process elements in the Milky Way and the higher occurrence rate of neutron star-white dwarf mergers, it is unlikely that such mergers can produce a significant amount of neutron-rich ejecta, with the generated mass of $r$-process elements being lower than $0.005M_{\odot}$.
\end{abstract}


\section{Introduction}
\label{sec:introduction}

The rapid neutron capture process ($r$-process) is considered as the primary mechanism responsible for the synthesis of heavy elements beyond iron in the universe \citep{1957RvMP...29..547B}. However, the astrophysical site where the $r$-process actually occurs remains an open question in astrophysics \citep{2021RvMP...93a5002C,2023A&ARv..31....1A}. For a long time, core-collapse supernovae (CCSNe) were regarded as the favored production site \citep{1994ApJ...433..229W,1996ApJ...471..331Q}, but subsequent research revealed that they are capable of producing only some of the lightest $r$-process nuclei \citep{2013JPhG...40a3201A,2013ApJ...770L..22W}. As an alternative site, the ejection of neutron-rich matter during the merger of binary neutron stars (BNS) or neutron star-black hole (NS-BH) system is considered an ideal site for $r$-process nucleosynthesis \citep{1974ApJ...192L.145L,1982ApL....22..143S}. Later studies suggested that such mergers would be accompanied by short-duration gamma-ray bursts (sGRBs, \citealp{1989Natur.340..126E,1992ApJ...395L..83N}) and kilonovae powered by the radioactive decay of freshly synthesized $r$-process nuclei \citep{1998ApJ...507L..59L,2010MNRAS.406.2650M,2012MNRAS.426.1940K,2013ApJ...774...25K,2013ApJ...775...18B,2016ApJ...829..110B,2019LRR....23....1M}. The subsequent radioactive decay of unstable nuclei in the merger ejecta during its optically thin phase may also produce a gamma-ray transient accompanying the kilonova \citep{2016MNRAS.459...35H,2019ApJ...872...19L,2021ApJ...919...59C,2022ApJ...932L...7C} and a MeV neutrino flash \citep{2023MNRAS.520.2806C}.

The first candidate kilonova was discovered in association with the sGRB 130603B, showing an excess of near-infrared emission consistent with the predictions of the kilonova emission model \citep{2013Natur.500..547T,2013ApJ...774L..23B,2013ApJ...779L..25F}. After that, several other sGRBs were claimed to be associated with kilonovae, e.g., GRB 050709 \citep{2016NatCo...712898J}, GRB 060614 \citep{2015NatCo...6.7323Y}, GRB 070809 \citep{2020NatAs...4...77J}, and GRB 080503 \citep{2015ApJ...807..163G}. The first direct evidence confirming the BNS merger as the source of kilonovae and sGRBs comes from the multi-messenger observation of the first neutron star merger event GW170817/GRB 170817A/AT2017gfo \citep{2017PhRvL.119p1101A,2017ApJ...848L..12A,2017PASA...34...69A,2017Natur.551...64A,2017Sci...358.1556C,2017ApJ...848L..17C,2017Sci...358.1570D,2017Sci...358.1565E,2017ApJ...848L..14G,2017Natur.551...80K,2017Sci...358.1559K,2017Natur.551...67P,2017Sci...358.1574S,2017Natur.551...75S,2017PASJ...69..102T,2017ApJ...848L..27T}. The luminosity and colour evolution of kilonova AT2017gfo suggest that approximately $0.05M_{\odot}$ of heavy $r$-process nuclei were synthesized in the merger ejecta \citep{2017Natur.551...80K,2024MNRAS.527.5540C}. Subsequent analysis of the observed spectrum further identified the presence of the heavy element strontium \citep{2019Natur.574..497W}.

Most recently, two kilonovae were discovered in association with long-duration gamma-ray bursts: GRB 211211A \citep{2022Natur.612..223R,2022Natur.612..228T,2022Natur.612..232Y} and GRB 230307A \citep{2023arXiv230702098L,2023arXiv230800633G,2023arXiv230800638Y}. The optical and near-infrared light curves of these two kilonovae exhibit very similar luminosities and decay rates compared to the kilonova AT2017gfo \citep{2022Natur.612..223R,2023arXiv230702098L}. The observed light curves indicate that these events produced an $r$-process ejecta with a mass of $\sim0.03-0.07M_{\odot}$ \citep{2022Natur.612..223R,2023arXiv230702098L}, consistent with estimated ejecta mass of kilonova AT2017gfo. The presence of near-infrared emission in the late phase suggests that a significant amount of lanthanide nuclei was synthesized in the merger ejecta \citep{2013ApJ...774...25K,2013ApJ...775...18B}. However, the duration of the prompt $\gamma$-ray emission in these events remains a challenge to explain.
\cite{2022Natur.612..232Y} proposed that the duration of GRB 211211A could be well explained by the merger of a neutron star-white dwarf (NS-WD) system. In their model, the kilonova associated with GRB 211211A was powered by the combined energy sources: spin-down energy from the central magnetar and radioactive energy from neutron-rich ejecta \citep{2013ApJ...776L..40Y,2018ApJ...861..114Y}. The ejecta opacity ($\kappa = 0.73$~g~cm$^{-2}$) obtained by \cite{2022Natur.612..232Y} is higher than the typical opacity of iron-group elements ($\kappa = 0.1$~g~cm$^{-2}$, \citealp{2013ApJ...774...25K}), generally supporting their assumption that NS-WD mergers can produce neutron-rich ejecta. However, the question of whether NS-WD mergers can indeed produce neutron-rich ejecta and generate heavy elements similar to BNS mergers remains unanswered. Recent hydrodynamic simulations have shown that the astrophysical conditions in NS-WD mergers are unsuitable for the production of heavy elements \citep{2019MNRAS.488..259F,2022MNRAS.510.3758B,2023ApJ...956...71K,2024A&A...681A..41M}. On the other hand, considering Galactic $r$-process observations, NS-WD mergers may pose significant challenges in the production of heavy elements. For example, if NS-WD mergers indeed contribute to heavy element production, the total mass of synthesized heavy elements may exceed the observed Galactic abundance of $r$-process elements. Consequently, the possibility of NS-WD mergers contributing to heavy element production through $r$-process nucleosynthesis can be constrained by the observation of Galactic $r$-process elements.

In this paper, our primary objective is to estimate the contribution of BNS mergers to the total mass $r$-process elements in the Milky Way and to investigate the potential of NS-WD mergers as alternative sites for $r$-process nucleosynthesis, based on an analysis of Galactic $r$-process observations. We employ an $r$-process nucleosynthesis network to examine heavy element production in BNS mergers, allowing us to evaluate their significance in the overall Galactic $r$-process enrichment. Given the observed similarities in luminosity and decline rate among kilonovae following GRB 211211A, GRB 230307A, and the well-studied kilonova AT2017gfo, we assume that the heavy element distribution in these events consistent with that of BNS mergers. We then estimate the allowed merger rate and heavy element yield for NS-WD mergers.
The paper is organized as follows. In Section~\ref{sec:method}, we provide details on $r$-process nucleosynthesis and describe the procedure used to calculate the contribution to the total mass of $r$-process elements in the Milky Way. Section~\ref{sec:results} presents the numerical results obtained from our analysis. The conclusion and discussion are presented in Section~\ref{sec:summary}.

\section{Method}
\label{sec:method}

\subsection{r-Process Nucleosynthesis}

We use the nuclear reaction network code SkyNet\citep{2015ApJ...815...82L,2017ApJS..233...18L} to perform $r$-process nucleosynthesis simulations for BNS mergers. SkyNet evolves the abundances of 7843 nuclide species and includes over 140,000 nuclear reactions. 
The nuclear reaction rates used in SkyNet are taken from the JINA REACLIB database \citep{2010ApJS..189..240C}. In our calculations, we use the latest nuclear database, including nuclear mass data from AME2020 \citep{2021ChPhC..45c0003W} and radioactive decay data from NUBASE2020 \citep{2021ChPhC..45c0001K}. For the nuclide species without experimental data, we use the theoretical data from nuclear physics models, including Finite-Range Droplet Model (FRDM2012; \citealp{2016ADNDT.109....1M}), Hartree-Fock-Bogoliubov (HFB27; \citealp{2009PhRvL.102o2503G}), Duflo-Zuker (DZ31; \citealp{1995PhRvC..52...23D}), and Weizs${\rm\ddot{a}}$cker-Skyrme (WS4; \citealp{2014PhLB..734..215W}). Neutron capture rates are calculated using the corresponding nuclear mass data through the nuclear reaction code TALYS \citep{2008A&A...487..767G}. Nuclear fission processes are calculated using fission barriers from \cite{2015PhRvC..91b4310M} and double Gaussian fission fragment distributions from \cite{1975NuPhA.239..489K}. 

The astrophysical trajectories for dynamical ejecta and wind ejecta are consistent with the numerical relativity simulations presented by \cite{2018ApJ...869..130R} and \cite{2021ApJ...906...98N}, respectively. We have selected simulation results calculated with the DD2 equation of state as the astrophysical inputs for our $r$-process nucleosynthesis calculations, as listed in Table~\ref{para}. We note that the choice of equation of state may have an impact on the simulation results of $r$-process nucleosynthesis for a single merger event. However, its impact on the calculation of the contribution to the total mass of $r$-process elements should be minimal, given that we will average the results from simulations of all individual merger events. Additionally, the neutrino treatment may also impact $r$-process nucleosynthesis in BNS mergers \citep{2020PhRvD.102j3015G,2021PhRvL.126y1101L,2023MNRAS.520.2806C}.

\begin{table}
    \centering
    \caption{Astrophysical inputs for $r$-process nucleosynthesis simulations in BNS mergers, encompassing the masses of neutron stars ($M_1$, $M_2$), the mass ratio of the binary system ($q$), electron fraction ($Y_{\rm e}$), entropy ($s$), and expansion velocity ($v$).}
    \begin{tabular}{cccccc}
    \hline\hline
    Model  &  $M_1+M_2$~($M_{\odot}$)  &  $q$  & 
 $Y_{\rm e}$  &  $s$~($k_{\rm B}$)  &  $v$~($c$)\\
    \hline
    M120120  &  1.20~+~1.20  &  1.00  &  0.15  &  21  &  0.14\\
    M125125  &  1.25~+~1.25  &  1.00  &  0.18  &  27  &  0.15\\
    M130130  &  1.30~+~1.30  &  1.00  &  0.13  &  15  &  0.18\\
    M135135  &  1.35~+~1.35  &  1.00  &  0.18  &  27  &  0.18\\
    M140140  &  1.40~+~1.40  &  1.00  &  0.17  &  22  &  0.22\\
    M150150  &  1.50~+~1.50  &  1.00  &  0.20  &  23  &  0.17\\
    M160160  &  1.60~+~1.60  &  1.00  &  0.14  &  13  &  0.24\\
    M140120  &  1.40~+~1.20  &  1.17  &  0.12  &  15  &  0.18\\
    M144139  &  1.44~+~1.39  &  1.04  &  0.17  &  22  &  0.20\\
    q100     &   --   &  1.00  &  0.30  &  10  &  0.17\\
    q120     &   --   &  1.20  &  0.26  &  10  &  0.15\\
    q143     &   --   &  1.43  &  0.26  &  10  &  0.16\\
    \hline\hline
    \end{tabular}
    \label{para}
\end{table}

\subsection{Contribution of BNS Mergers}

Based on the $r$-process nucleosynthesis simulation of BNS mergers, the mass fraction of each nuclide can be determined as
\begin{equation}
    X_i = A_i Y_i,
\end{equation}
where $A_i$ is the atomic mass number of the $i$th nuclide and $Y_i$ is the corresponding abundance. The contribution of BNS mergers to the total mass of $r$-process elements in the Milky Way is given by \citep{2015NatPh..11.1042H,2018IJMPD..2742005H,2019EPJA...55..203S}
\begin{equation}
    M_i \approx T_{\rm MW} \mathcal{R}_{\rm MW} X_i M_{\rm ej},
\end{equation}
where $T_{\rm MW} \approx 10$~Gyr is the age of the Milky Way, $\mathcal{R}_{\rm MW}$ is the local merger rate, and $M_{\rm ej}$ is the mass value of ejected materials. Assuming a local density of Milky-Way equivalent galaxies to be  $\rho_{\rm gal}=1.16\times10^{-2}$~Mpc$^{-3}$ \citep{2008ApJ...675.1459K,2010CQGra..27q3001A}, the local merger rate $\mathcal{R}_{\rm MW}$, in units of Myr$^{-1}$, can be converted to the volumetric event rate $R$ in Gpc$^{-3}$~yr$^{-1}$ via
\begin{equation}
    R = 10^{3}\rho_{\rm gal} \mathcal{R}_{\rm MW}~{\rm Gpc^{-3}~yr^{-1}}.
\end{equation}
The total mass of $r$-process elements can be obtained by summarizing the contributions of all nuclide species:
\begin{equation}
    M_{\rm tot} = \sum_i M_i.
\end{equation}

In our calculations, we employ the SkyNet network to calculate the abundance of heavy elements for a series of individual BNS merger events. The heavy element abundances from these individual merger events are then averaged to estimate their collective contribution to Galactic $r$-process elements. We adopt the local event rate of BNS mergers inferred from recent LIGO/Virgo observations ($R_{\rm BNS}=320^{+490}_{-240}$~Gpc$^{-3}$~yr$^{-1}$, \citealp{2021PhRvX..11b1053A}). The mass value of ejected materials from BNS mergers is set as $M_{\rm ej}=0.01_{-0.005}^{+0.04} M_{\odot}$. It is worth noting that some BNS mergers may not contribute to Galactic $r$-process elements since they can escape the Galactic disk due to natal kicks \citep{2018IJMPD..2742005H,2021MNRAS.505.5862W}. Additionally, our calculation does not consider mass loss caused by the Galactic wind. These two effects may lead to an overestimation of the contribution from BNS mergers. Therefore, accounting for these effects may impact the overall contribution of BNS mergers to the Galactic $r$-process elements.

\subsection{Constraints on Heavy Element Yield in NS-WD Mergers}

To estimate the heavy element yield of NS-WD mergers, we assume that the primary contribution to the total mass of heavy elements in the Milky Way comes from BNS mergers. The NS-WD mergers serve as an additional source for heavy element production, contributing to the remaining heavy elements in our galaxy. The total stellar mass of the Milky Way is $6.43 \times 10^{10} M_{\odot}$ \citep{2011MNRAS.414.2446M}, and the estimated total mass of heavy $r$-process elements in the Milky Way is $5.7\times10^{3} M_{\odot}$. Following \cite{2018A&A...619A..53T}, the event rate of NS-WD mergers falls within the range of $(3.7 - 18) \times 10^{-5}M_{\odot}^{-1}$. Assuming a constant star formation rate of $3M_{\odot}/{\rm yr}$, the NS-WD merger rate ranges from $111$ to $540$ Myr$^{-1}$, consistent with the rate of $260$ Myr$^{-1}$ based on binary radio pulsars \citep{2017MNRAS.467.3556B}. To account for uncertainties in the merger rate and ejecta mass of BNS mergers, we consider the minimum contributions from BNS events (i.e., setting the BNS merger rate to $80$ Gpc$^{-3}$ yr$^{-1}$ and the ejecta mass to $0.005 M_{\odot}$) to derive an upper limit on the heavy element yield from NS-WD mergers.

\section{Results}
\label{sec:results}

\begin{figure}
    \centering
    \includegraphics[width=0.55\textwidth]{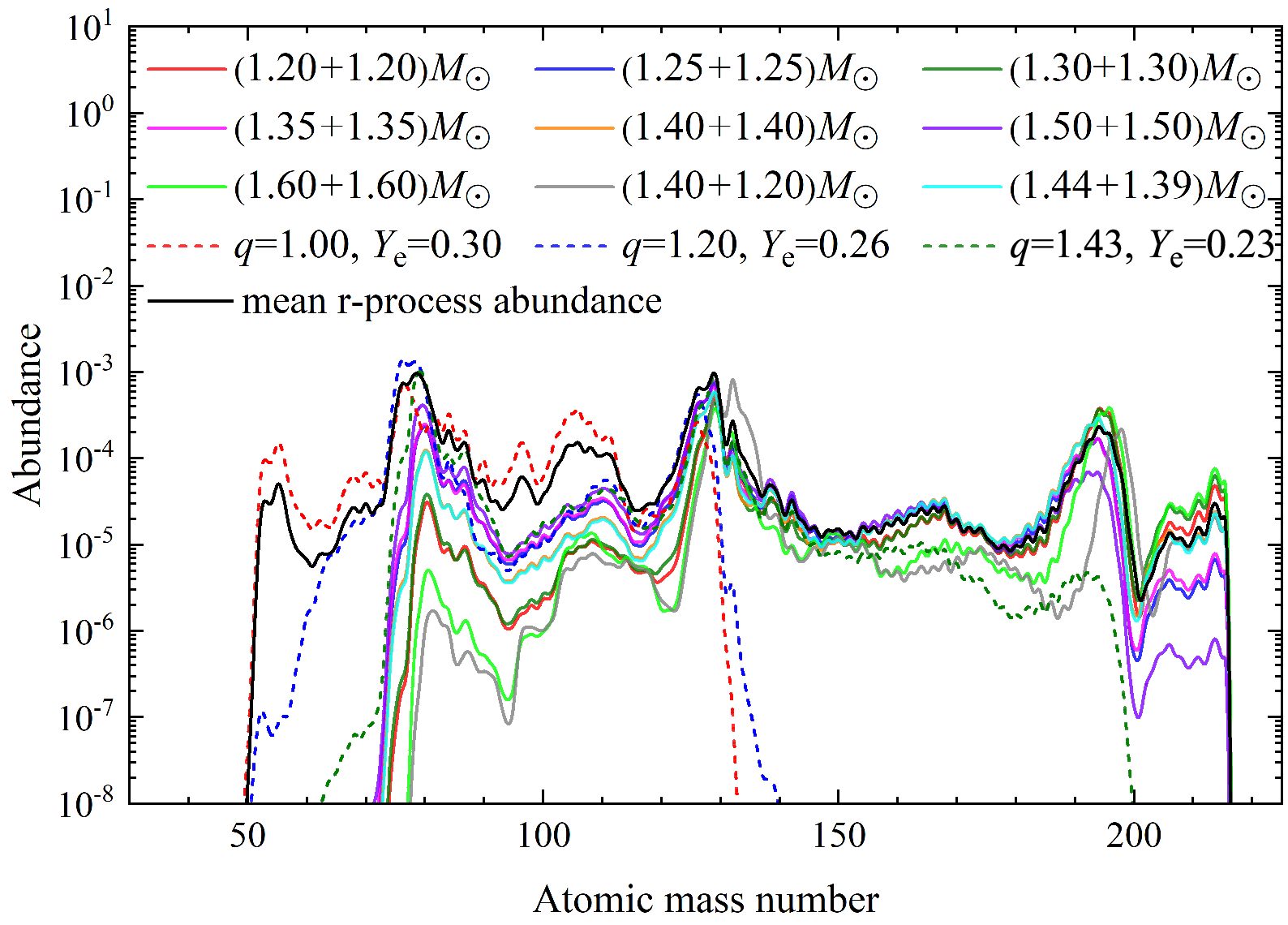}
    \caption{The resulting $r$-process abundances for dynamical ejecta (solid lines) and disk wind ejecta (dashed lines) in BNS mergers. Astrophysical inputs for $r$-process nucleosynthesis simulations are listed in Table~\ref{para}, encompassing the mass ratio of binary system ($q$) and the electron fraction ($Y_{\rm e}$).}
    \label{abundance}
\end{figure}

Figure~\ref{abundance} shows the resulting abundances from $r$-process nucleosynthesis simulations of BNS mergers. The dynamical ejecta is sufficiently neutron rich, leading to the production of a broad range of heavy elements spanning from $A=70$ to $A=220$. The resulting $r$-process abundances exhibit three characteristic peaks at $A\sim80, \sim130, \sim195$, respectively, consistent with the observed solar $r$-process abundance pattern \citep{2007PhR...450...97A}. Although the $r$-process abundances exhibit some variations with different neutron star masses, the positions of the three characteristic peaks are consistently reproduced. In the case of wind ejecta, the ejected materials lack the necessary neutron richness to produce a substantial amount of heavy $r$-process elements. Instead, the wind ejecta primarily contributes to the production of light $r$-process elements, particularly nuclei within the range of $A=50$ to $A=130$.

\begin{figure}
    \centering
    \includegraphics[width=0.55\textwidth]{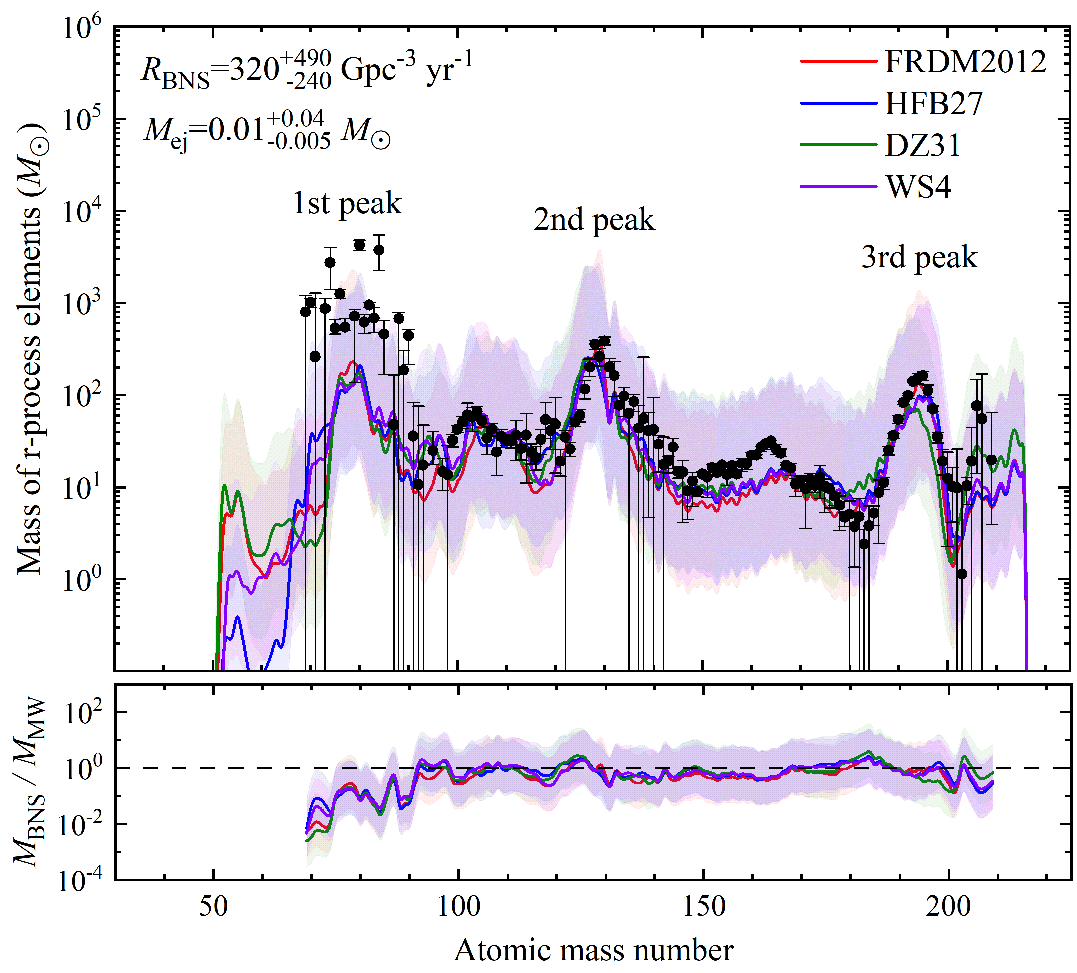}
    \caption{Top panel: Contribution of BNS mergers to total mass of $r$-process elements in the Milky Way. Simulations of $r$-process nucleosynthesis are conducted using four widely-used nuclear physics models: FRDM2012 \citep{2016ADNDT.109....1M}, HFB27 \citep{2009PhRvL.102o2503G}, DZ31 \citep{1995PhRvC..52...23D}, and WS4 \citep{2014PhLB..734..215W}. We adopt the local event rate of BNS mergers inferred from recent LIGO/Virgo observations ($R_{\rm BNS}=320^{+490}_{-240}$~Gpc$^{-3}$~yr$^{-1}$, \citealp{2021PhRvX..11b1053A}). The mass value of ejected materials from BNS mergers is set as $M_{\rm ej}=0.01_{-0.005}^{+0.04} M_{\odot}$. The shaded region represents the uncertainty arising from variations in the merger rate and ejecta mass. The total stellar mass of the Milky Way is set to be $6.43 \times 10^{10} M_{\odot}$ \citep{2011MNRAS.414.2446M}. The black dots represent the solar system $r$-process abundances taken from \protect\cite{2007PhR...450...97A}. Bottom panel: Deviations in the mass of $r$-process elements calculated by four nuclear physics models with respect to the total mass of $r$-process elements in the Milky Way.}
    \label{Mass_BNS}
\end{figure}

Based on detailed $r$-process nucleosynthesis simulations for individual BNS merger events, we estimate their contribution to the total mass of $r$-process elements in the Milky Way, as shown in Figure~\ref{Mass_BNS}. For comparison, black dots represent the solar system $r$-process abundances obtained from \cite{2007PhR...450...97A}. It is found that the contribution from BNS mergers is in agreement with Galactic $r$-process elements, particularly for heavy $r$-process elements with $A\geq90$. The second ($A\sim130$) and third ($A\sim195$) $r$-process peak features in the observed data are well reproduced. This remarkable consistency suggests that BNS mergers are the dominant contributors to the production of heavy $r$-process elements. The total mass of heavy elements generated by BNS merger events ranges from $3.45\times10^2M_\odot$ to $3.49\times10^4M_\odot$. However, in the region of light $r$-process elements ($69 \leq A \leq 90$), the contribution from BNS mergers is not sufficient to fully account for the observed $r$-process data. This result implies that other astrophysical sites, such as CCSNe, should play a significant role in the production of light $r$-process elements.

\begin{figure}
    \centering
    \includegraphics[width=0.55\textwidth]{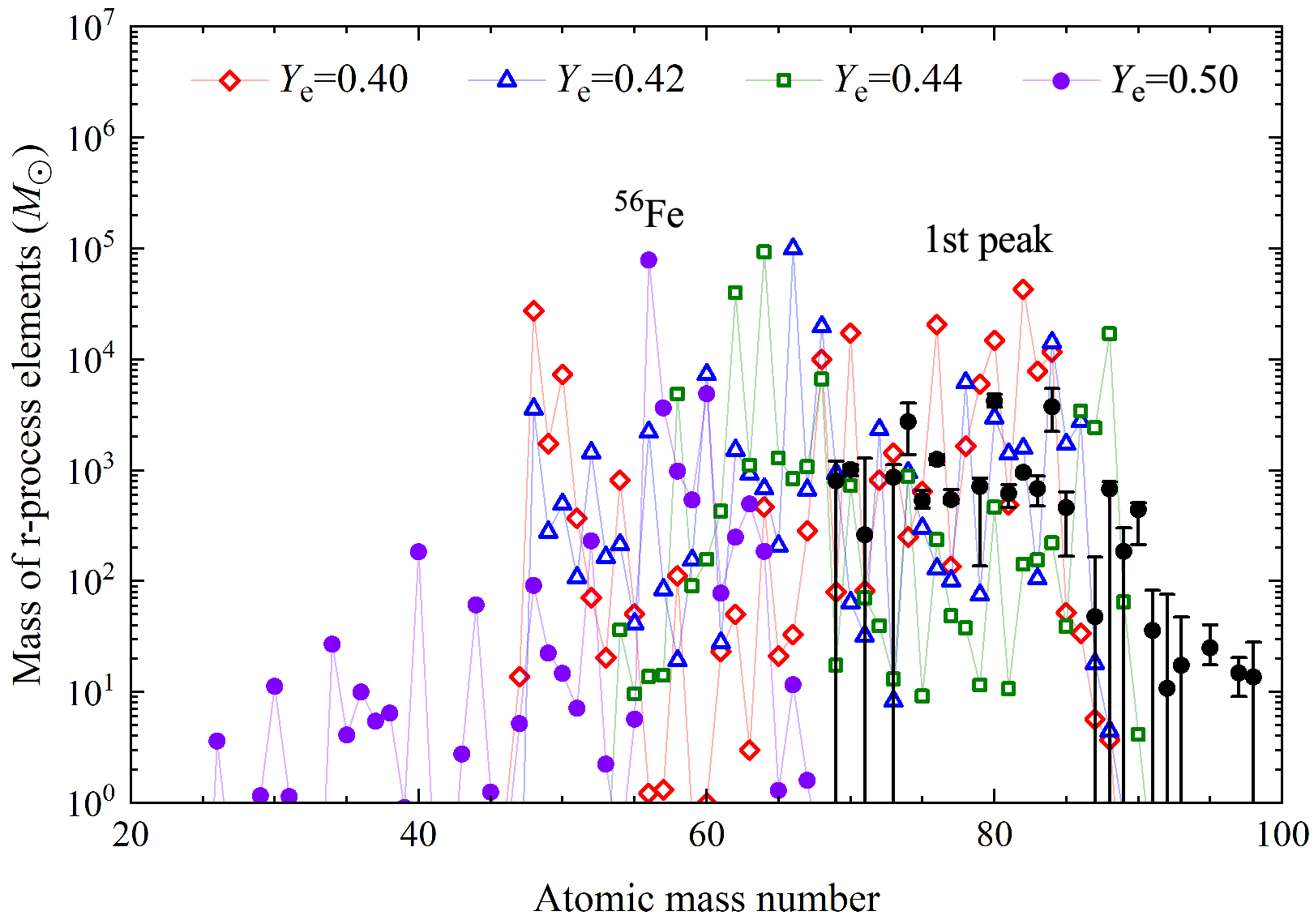}
    \caption{Contribution of CCSNe to the total mass of $r$-process elements in the Milky Way. We assume that $25\%$ of CCSNe with low-mass progenitors ($\leq9.6M_{\odot}$) can produce neutron-rich ejecta, with a typical mass value of $0.01M_{\odot}$ \citep{2018ApJ...852...40W,2023ApJ...954..114W}. The local event rate of CCSNe is set to $7.05_{-1.25}^{+1.43} \times 10^4$~Gpc$^{-3}$~yr$^{-1}$ \citep{2011MNRAS.412.1473L}.}
    \label{Mass_CCSN}
\end{figure}

To estimate the contribution of CCSNe to total mass of $r$-process elements in the Milky Way, we conducted $r$-process nucleosynthesis simulations for CCSNe using the SkyNet network. The astrophysical inputs for our simulations are generally consistent with recent three-dimensional CCSN simulations presented by \cite{2023ApJ...954..114W}, featuring an entropy of $s=30$~$k_{\rm B}$/baryon and an expansion timescale of $\tau=30$~ms. \cite{2018ApJ...852...40W} shown that the neutron-rich ejecta tend to be ejected in CCSNe from low-mass progenitors, specifically those with masses less than $9.6M_{\odot}$. According to the initial mass function $dN/dM \propto M^{-2.3}$, the fraction of CCSNe with progenitor masses less than $9.6M_{\odot}$ in the mass range from $8M_{\odot}$ to $40M_{\odot}$ is $\sim25\%$ \citep{2003ApJ...591..288H}. Therefore, we assume that $\sim25\%$ of the ejected matter from CCSNe is neutron-rich, corresponding to an electron fraction $Y_{\rm e}=0.40\sim0.50$. Figure~\ref{Mass_CCSN} shows the contribution of CCSNe to the total mass of $r$-process elements in the Milky Way. In scenarios where the electron fraction $Y_{\rm e}=0.50$, the wind matter can only produce nuclei around the iron peak and does not contribute to $r$-process elements. In the case of  $Y_{\rm e}<0.50$, the wind matter from CCSNe can generate light $r$-process elements around the first $r$-process peak ($A\sim80$). As can be seen from Figure~\ref{Mass_CCSN}, the contribution of CCSNe to the first $r$-process elements is broadly consistent with the solar system $r$-process abundance. This suggests that some fractions of CCSNe are the primary source for the production of light $r$-process elements in the universe. Note that we have only provided a rough estimate of the contribution of CCSNe to light $r$-process elements, and detailed $r$-process nucleosynthesis calculations involving various mass progenitors of CCSNe are required for further confirmation.

\begin{figure}
    \centering
    \includegraphics[width=0.55\textwidth]{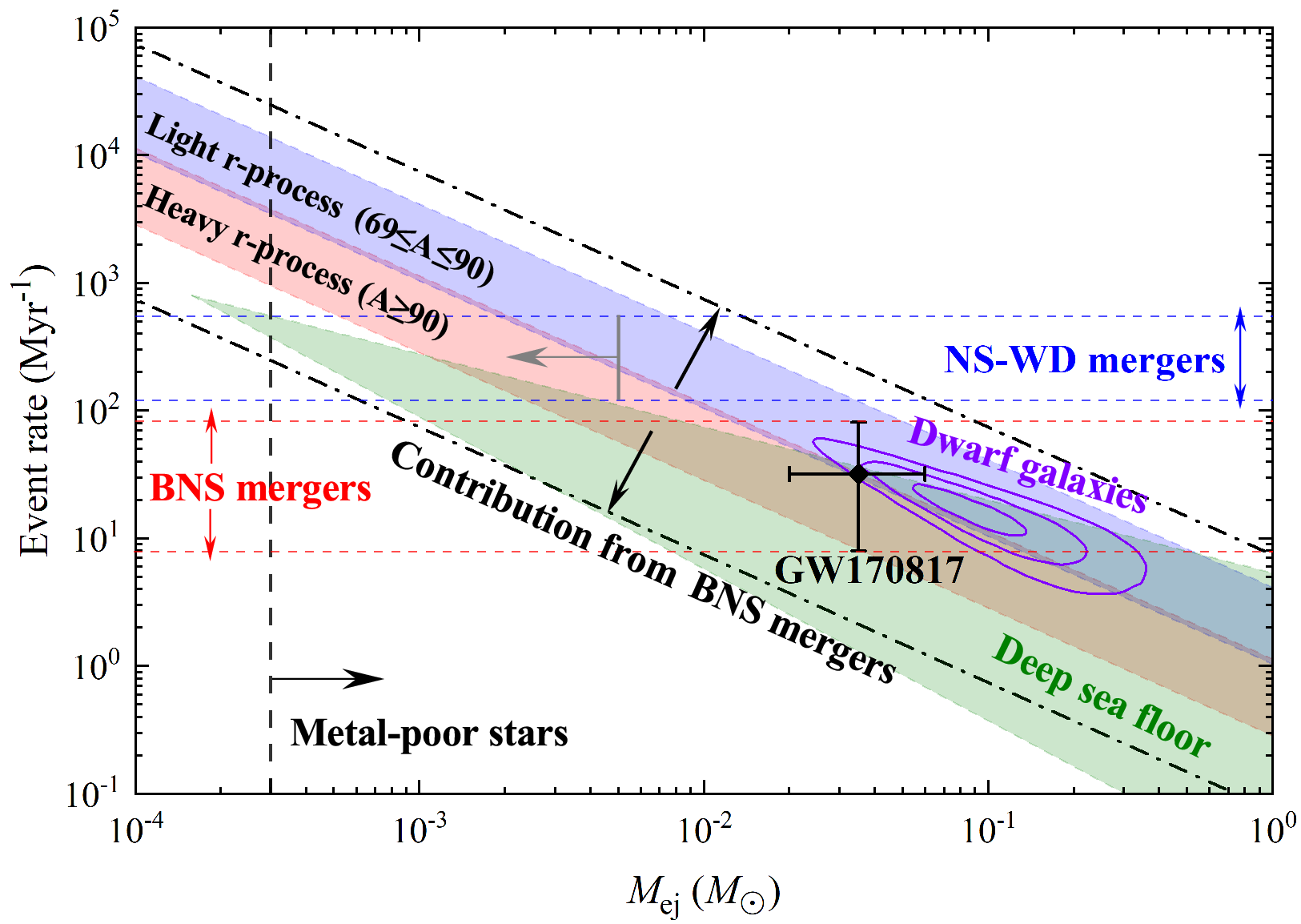}
    \caption{Constraints on the local event rate and heavy element yield per event. The red and blue shaded regions represent the total mass of heavy $r$-process elements ($A \geq 90$) and light $r$-process elements ($69 \leq A \leq 90$), respectively. The green shaded region corresponds to the constraint on $r$-process enrichment derived from the measurement of $^{244}$Pu in the deep sea floor \citep{2015NatPh..11.1042H}. The purple contours represent the constraints inferred from the observation of Eu element in dwarf galaxies \citep{2016ApJ...832..149B}. The black dashed line indicates the lower limit on ejecta mass per event, as constrained by the analysis of metal-poor stars \citep{2018ApJ...860...89M}. The red and blue dashed lines represent the event rates of BNS mergers and NS-WD mergers, respectively. The gray line indicates a mass upper limit of $0.005M_{\odot}$ for the heavy element yield in NS-WD mergers.}
    \label{rate}
\end{figure}

Based on observations of Galactic $r$-process elements, we can constrain the local event rate and the yield of heavy element per event, as shown in Figure~\ref{rate}. The red and blue shaded regions in the figure represent the total mass of heavy and light $r$-process elements, respectively. It can be observed that the contribution of BNS mergers is not only consistent with the total mass of $r$-process elements in the Milky Way but also matches the constraints on the local event rate and the yield of heavy element per event. These constraints are derived from various measurements, including those of $^{244}$Pu in the deep sea floor and the observation of the Eu element in dwarf galaxies. The derived merger rate and ejecta mass from the BNS merger event GW170817 also agree with theses constraints. Furthermore, the contribution of BNS mergers meets the lower limit on the ejecta mass constrained by the analysis of metal-poor halo stars. These results provide further confirmation of the dominant role played by BNS mergers in the production of heavy $r$-process elements. 

The potential contribution of NS-WD mergers to the total mass of $r$-process elements in the Milky Way is also shown in Figure~\ref{rate}. Given the similarities in luminosity and decline rate between the kilonovae following GRB 211211A and GRB 230307A and the kilonova AT2017gfo resulting from a BNS merger event, we assume that the heavy element distributions of these events are consistent with those of BNS mergers. Considering that the event rate of NS-WD mergers is approximately an order of magnitude higher than that of BNS mergers \citep{2018A&A...619A..53T}, the mass of neutron-rich ejecta produced by NS-WD mergers should be correspondingly smaller by about an order of magnitude compared to BNS mergers. Otherwise, it could lead to an overestimation of the contribution of such events to the total mass of $r$-process elements in the Milky Way. Taking into account the uncertainties in the merger rate and ejecta mass of BNS systems, if we assume a scenario where BNS mergers contribute maximally to the total mass of heavy elements in the Milky Way (setting the BNS merger rate to $810$ Gpc$^{-3}$ yr$^{-1}$ and the ejecta mass to $0.05 M_{\odot}$), then the minimum ejecta mass from NS-WD mergers could be zero. Conversely, if BNS mergers contribute minimally to the total mass of heavy elements (setting the BNS merger rate to $80$ Gpc$^{-3}$ yr$^{-1}$ and the ejecta mass to $0.005 M_{\odot}$), the maximum ejecta mass from NS-WD mergers could be $0.005M_{\odot}$. In other words, the analysis of the total mass of heavy elements in the Milky Way suggests that the maximum allowed yield for heavy $r$-process elements from NS-WD mergers is $0.005M_{\odot}$, with the corresponding maximum allowed yield for lanthanide elements being $5\times 10^{-4}M_{\odot}$.

\section{Summary and Discussion}
\label{sec:summary}

The recent identification of kilonovae following long-duration gamma-ray bursts GRB 211211A and GRB 230307A has triggered discussions about the possibility of neutron star-white dwarf (NS-WD) mergers to produce significant amounts of neutron-rich ejecta and generate $r$-process elements. While the duration of the prompt $\gamma$-ray emission in these events can be explained by the merger of a NS-WD system, the question of whether NS-WD mergers can indeed generate neutron-rich ejecta comparable to binary neutron star (BNS) mergers remains a topic of ongoing research. In this paper, we have investigated the potential of NS-WD mergers in the production of heavy elements from the perspective of Galactic $r$-process observations. 

We first assess the contribution of BNS mergers to the Galactic $r$-process production. In contrast to previous studies that primarily focused on comparing the general features of $r$-process abundance patterns resulting from BNS mergers with those of solar $r$-process residuals, we adopt a quantitative approach to calculate the actual impact of BNS mergers on the production of heavy elements within the Milky Way. We use the nuclear reaction network SkyNet, updated with the latest nuclear database, to simulate a series of individual BNS merger events and determine the abundance of heavy elements (Figure~\ref{abundance}). By averaging the abundance results from these merger events, we estimate the collective contribution of BNS mergers to the total mass of $r$-process elements in the Milky Way. Our analysis reveals that the contribution of BNS mergers is consistent with the total mass of $r$-process elements in the Milky Way, especially for heavy $r$-process elements with atomic mass number $A\geq90$ (Figure~\ref{Mass_BNS}). The total mass of heavy elements generated by BNS merger events ranges from $3.45\times10^2M_\odot$ to $3.49\times10^4M_\odot$. Furthermore, the derived merger rate and ejecta mass of the BNS merger event GW170817 are in good agreement with constraints on $r$-process enrichment inferred from the measurement of $^{244}$Pu in the deep sea floor \citep{2015NatPh..11.1042H} and the observation of the Eu element in dwarf galaxies \citep{2016ApJ...832..149B}. Additionally, the ejecta mass of merger event GW170817 meets the lower limit on the yield of $r$-process material per event derived from analyses of metal-poor halo stars. These results suggest that BNS mergers are the dominant contributors to the production of heavy $r$-process elements. 

We note that although BNS mergers can indeed provide an explanation for the origin of heavy $r$-process elements in the Milky Way, they can not fully account for all $r$-process elements. For the production of light $r$-process elements ($69\leq A \leq90$), it is possible that core-collapse supernovae (CCSNe) play a more significant role compared to BNS mergers (Figure~\ref{Mass_CCSN}). In other words, heavy elements originate from both BNS mergers and CCSNe, with BNS mergers primarily contributing to the heavy $r$-process elements. This result is consistent with observations of metal-poor halo stars, where the abundances of light $r$-process elements deviate from solar-system $r$-process abundances \citep{2008ARA&A..46..241S}, suggesting diverse production sites for these elements \citep{2004ApJ...601..864T,2019ApJ...885...33H}.

Given the similarities in luminosity and decline rate between the kilonovae following GRB 211211A and GRB 230307A and the kilonova AT2017gfo resulting from a BNS merger event, we estimate the potential contribution of NS-WD mergers by assuming that heavy element distributions in these events are consistent with that of BNS mergers. Due to the higher event rates of NS-WD mergers, the allowed mass of neutron-rich ejecta should be lower than $0.005M_{\odot}$ through the analysis of Galactic $r$-process elements. This result suggests that the NS-WD mergers can not produce significant amounts of neutron-rich ejecta compared to BNS mergers. \cite{2022Natur.612..232Y} proposed that the mass of neutron-rich ejecta obtained by fitting the kilonova light curve associated with GRB 211211A is $0.037_{-0.004}^{+0.008}M_{\odot}$, significantly exceeding the allowed upper limit for heavy element yield from NS-WD mergers. Notably, recent simulations have shown that the material ejected from NS-WD mergers is not sufficiently neutron-rich to generate heavy $r$-process nuclei \citep{2019MNRAS.488..259F,2022MNRAS.510.3758B,2023ApJ...956...71K,2024A&A...681A..41M}. If the kilonova associated with GRB 211211A is driven by NS-WD merger, then the energy source of the kilonova light curve should not be originating from heavy $r$-process elements. \cite{2023ApJ...947L..21Z} proposed that the kilonova following GRB 211211A may be powered by the combined energy sources: the radioactive decay energy from $^{56}$Ni and spin-down energy from the post-merger magnetar.

It is worth noting that, besides BNS mergers, neutron star-black hole (NS-BH) mergers are also considered potential astrophysical sites for the production of heavy $r$-process elements. However, the capability of NS-BH mergers to produce heavy elements remains uncertain, as no observational evidence of kilonovae associated with NS-BH mergers has been discovered so far. On the other hand, the local event rate of NS-BH mergers is estimated to be $R_{\rm NSBH} = 45^{+75}_{-33}$~Gpc$^{-3}$~yr$^{-1}$ \citep{2021ApJ...915L...5A}, which is significant lower than that of BNS mergers. As a result, although NS-BH mergers are unlikely to be the dominant contributors to the production of heavy $r$-process elements, they could play a crucial role in complementing the overall production of heavy elements.

Another possible astrophysical sites for the production of heavy $r$-process elements are collapsars, which are formed from the collapse of rapidly rotating massive stars. \cite{2019Natur.569..241S} proposed that disk wind ejecta from collapsars can attain the required neutron-rich conditions to synthesize both light and heavy $r$-process elements through $r$-process nucleosynthesis. However, it is important to note that recent simulations adopting different methods of neutrino transport do not consistently observe the production of the heavy $r$-process nuclei \citep{2020PhRvD.102l3014F,2020ApJ...902...66M,2022ApJ...934L..30J}. Moreover, the late-time observation of the nearby gamma-ray burst, GRB 221009A (originating from the collapse of a rapidly rotating massive star), conducted by the James Webb Space Telescope, did not find evidence of $r$-process nucleosynthesis \citep{2023arXiv230814197B}.

\section*{Acknowledgements}

We thank Jin-Ping Zhu and Ning Wang for fruitful discussion.
This work was supported by the National Natural Science Foundation of China (Grant Nos.~11973014, 12133003, and 12347172) and the National Funded Postdoctoral Program of China (Grant No.~GZB20230029).

\begin{center}
{\bf ORCID iDs}
\end{center}

Meng-Hua Chen: https://orcid.org/0000-0001-8406-8683

Li-Xin Li: https://orcid.org/0000-0002-8466-321X

Qiu-Hong Chen: https://orcid.org/0009-0006-8625-5283

Rui-Chong Hu: https://orcid.org/0000-0002-6442-7850

En-Wei Liang: https://orcid.org/0000-0002-7044-733X

\section*{Data Availability}
 
The data underlying this paper can be shared on reasonable request to the corresponding author.


\newpage
\onecolumn

\bsp	
\label{lastpage}
\end{document}